\documentclass[fleqn,12pt]{wlscirep}

\usepackage{color,soul}
\usepackage{upgreek}
\usepackage{mathastext}
\usepackage{setspace}
\title{\centering Wave-based extreme deep learning based on non-linear time-Floquet entanglement}

\author[1]{Ali Momeni}
\author[1,*]{Romain Fleury}
\affil[1]{Laboratory of Wave Engineering, School of Electrical Engineering, Swiss Federal Institute of Technology in Lausanne (EPFL), Lausanne, Switzerland}

\affil[*]{E-mail: romain.fleury@epfl.ch}

\doublespace 

\begin{abstract}
Wave-based analog signal processing holds the promise of extremely fast, on-the-fly, power-efficient data processing, occurring as a wave propagates through an artificially engineered medium. Yet, due to the fundamentally weak non-linearities of traditional wave materials, such analog processors have been so far largely confined to simple linear projections such as image edge detection or matrix multiplications. Complex neuromorphic computing tasks, which inherently require strong non-linearities, have so far remained out-of-reach of wave-based solutions, with a few attempts that implemented non-linearities on the digital front, or used weak and inflexible non-linear sensors, restraining the learning performance.
Here, we tackle this issue by demonstrating the relevance of Time-Floquet physics to induce a strong non-linear entanglement between signal inputs at different frequencies, enabling a power-efficient and versatile wave platform for analog extreme deep learning involving a single, uniformly modulated dielectric layer and a scattering medium. We prove the efficiency of the method for extreme learning machines and reservoir computing to solve a range of challenging learning tasks, from forecasting chaotic time series to the simultaneous classification of distinct datasets. Our results open the way for wave-based machine learning with high energy efficiency, speed and scalability.
\end{abstract}
\begin{document}

\flushbottom
\maketitle
\thispagestyle{empty}

\section{Introduction}
Recently, artificial intelligence (AI) systems based on advanced machine learning algorithms have attracted a surge of interest for their potential applications in processing the information hidden in large datasets  \cite{lecun2015deep,shen2017deep}. Wave-based analog implementations of these schemes, exploiting microwave or optical neural networks, promise to revolutionize our ability to perform a large variety of challenging data processing tasks by allowing for power-efficient and fast neuromorphic computing at the speed of light. Indeed, wave-based analog processors work directly in the native domain of an analog signal, processing it while the wave propagates through an engineered artificial structure (metamaterials and metasurfaces) \cite{engheta2006metamaterials,achouri2021electromagnetic,li2019machine,lin2014dielectric,momeni2018information,kiani2020self,kiani2020spatial,hosseininejad2019digital,rajabalipanah2019asymmetric,hosseininejad2019reprogrammable}, as previously established in the cases of simple linear operations such as image differentiation, signal integration and integro-differential equations solving\cite{silva2014performing,camacho2021single,estakhri2019inverse,zangeneh2020analogue,zangeneh2019topological,momeni2019generalized,babaee2021parallel, momeni2021switchable,momeni2020reciprocal,momeni2021asymmetric,abdolali2019parallel}. For more complex processing tasks, for example image recognition or speech processing, both non-linearity and a high degree of interconnection between the elements are desired, requirements that have led to various proposals of neuromorphic processors exploiting optical diffraction, coupled waveguide networks, or disordered structures \cite{wetzstein2020inference,lin2018all, zhang2021optical,zuo2019all,qian2020performing,xu2020photonic,   feldmann2019all,hughes2019wave,hamerly2019large,bueno2018reinforcement}. A particularly vexing challenge, however, is the implementation of non-linear processing elements. While power-efficient neuromorphic schemes require a pronounced, particular form of non-linearities, optical non-linearities, such as in Kerr dielectrics,  are typically weak at low intensitites, and cannot be much controlled. This leads to sub-optimal systems that must operate with high input powers\cite{hughes2019wave,papp2020nanoscale,teugin2020scalable}. As an alternative, non-linearities that are external to the wave-based processor have also been considered, for example by exploiting the intensity dependency of a sensor, that needs an additional electronic interconnection. Unfortunately, exploiting such weak and non-controllable non-linearities drastically confine the performance of most machine learning schemes, and the relevance of wave-base platforms have so far been largely restricted to the implementation of simple linear matrix projections.

Here, we propose to leverage the physics of wave systems that are periodically modulated in time, the so-called time-Floquet systems\cite{fleury2016floquet, wang2020nonreciprocity, koutserimpas2018nonreciprocal, estep2014magnetic,hadad2015space,koutserimpas2018parametric}, to solve this vexing challenge by implementing a strong, controllable non-linear entanglement between all the input signals. We propose to use a simple, thin, uniform dielectric slab, whose refractive index is slowly and weakly modulated in time. With the addition of linear random scattering disorder, we implement very efficient recurrent neural networks (RNNs) schemes, namely extreme learning machine (ELM) and reservoir computing (RC). We demonstrate the high accuracy of our Floquet extreme learning machine in challenging computing tasks, from the processing of one-dimensional data (learning non-linear functions), to challenging multi-dimensional data (e.g. the abalone dataset classification problem). We also demonstrate the flexibility of our scheme that can be multiplexed to tackle two unrelated classification tasks at the same time, simultaneously sorting COVID-19 Xray \begin{figure*}[!ht]
\centering
\includegraphics[width=\linewidth]{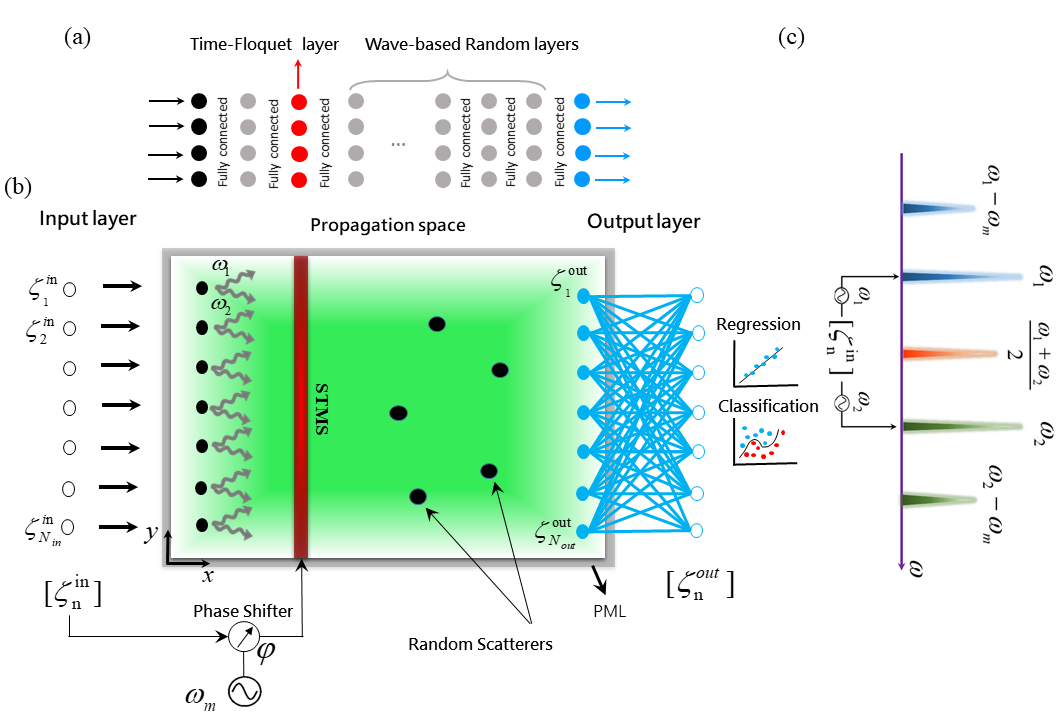}
\caption{\textbf{Wave based  time-Floquet extreme learning machine.} (a) Schematic of a neural network including a time-Floquet layer made from neurons whose properties are modulated periodically in time, and traditional random layers. Only the last layer (output) is trainable.
(b)  Concrete implementation with electromagnetic waves. The input signals $\zeta^{in}_n$ are modulated at $\omega_1$ and $\omega_2$. Their sum forms input signals that are independently radiated into the surrounding space by an array of source antennas (black disks). As the waves propagates in the green region,  they encounter a thin dielectric slab whose index of refraction is modulated at the frequency $\omega_m=|\omega_1-\omega_2|/2$, as well as five sub-wavelength scatterers, randomly located in the domain. The modulation phase is assumed to depend on the input vector $\zeta^{in}_n(t)$. The  gray rectangle represents an absorbing boundary layer. The output $\zeta^{out}_n(t)$ are fed into an adaptable blue dense layer, and used for regression and classification. (c) Non-linear phase entanglement. The modulated slab mixes signals at $\omega_1$ and $\omega_2$ into Floquet Hamonics spaced by $\omega_m$, whose amplitudes depend non-linearly on the input vector.      }
\label{Fig_1}
\end{figure*}lung images and hand-written digits. Finally, we validate our Floquet RC by predicting the time-evolution of a chaotic system over a large time-period (the Mackey-Glass time series). Such extreme time-Floquet analog learning machines are not only fast, easy-to-train, power efficient, and versatile, but also feature a unique accuracy performance that is comparable to that obtained with the best digital schemes.

\section{Nonlinear time-Floquet-based extreme learning machine }

We consider a particular class of neural networks, known as recurrent neural networks (RNNs). RNNs are ideal to process intricate data due to the internal cyclic connections between internal neurons, whose outputs depend on both the current inputs and the previous states of the neurons \cite{moon2019temporal}. This  memory effect allows RNNs to detect recursive relations in the data, which is relevant for example to process temporal signals. In digital implementations, however, the heavy internal connectivity matrices that are involved in the training process make RNNs particularly computationally-expensive and complicated \cite{nakajima2021scalable,du2017reservoir,zhong2021dynamic,midya2019reservoir}. In order to solve these challenges, a number of alternative computing approaches such as long short-term memory (LSTM) \cite{hochreiter1997long},  echo state networks (ESNs) \cite{dong2018scaling}, extreme learning machines (ELMs) \cite{huang2006extreme,marcucci2020theory,pierangeli2021photonic}, and reservoir computing (RC) \cite{vandoorne2014experimental,du2017reservoir,zhong2021dynamic,midya2019reservoir} have emerged. These schemes are particularly well suited for wave-based implementations, because waves propagation inherently relies on the inertial memory of the medium, which can be enhanced and engineered by leveraging resonant cavities, or multiple scattering. In addition, wave interferences are a particularly efficient way to create a high degree of interconnections between a large set of inputs.

Our time-Floquet neuromorphic processor implements an ELM, schematically shown in \textcolor{blue}{Fig. \ref{Fig_1}(a)}. ELMs, or closely related methods based on random neural networks \cite{pao1994learning}or support vector machines \cite{suykens1999least}, are a powerful scheme in which only a last layer of connections is trained (in blue). The fundamental mechanism is the use of the non-trained part of the network, whose layers are represented in grey and red in \textcolor{blue}{Fig. \ref{Fig_1}(a)}, to establish a nonlinear mapping between the initial space of the dataset, and a higher-dimensional feature space, where a properly trained classifier performs the separation and classification. In our case, this non-linear mapping is performed by letting one of the non-trained layers (in red) be weakly modulated in time at a frequency much lower than the one of the signal, but with a phase that depends on the input state.

%Similar to ELMs, RC uses a dynamic reservoir with fixed internal connections to induce a nonlinear mapping of inputs into a higher-dimensional feature space, which causes the initial inputs to become linearly separable in this new space \cite{zhong2021dynamic}. The reservoir should have some degree of nonlinearity in its dynamics, and unlike ELMs, some form of “fading memory”, meaning that it will gradually forget previous inputs as new inputs come in \cite{tanaka2019recent}.

A concrete implementation of this scheme in a wave platform is shown in \textcolor{blue}{Fig. \ref{Fig_1}(b)}. It consists of three parts:  
(i), an array of monopole antennas that radiates the various components of the input vector into the surrounding medium; (ii), a propagation space composed of a few scatterers and a thin dielectric slab, called a scattering time-modulated slab (STMS), whose index of refraction is weakly modulated in time; and (iii), the output layer made of an array of receiving antennas and a single  dense layer, digitally trained to perform the desired regression or classification tasks. At the input layer, the input vector $\zeta^{\text{in}}$ with components  $\zeta^{\text{in}}_1$, ..., $\zeta^\text{{in}}_N$ is first encoded into $N$ signals $s^{\text{in}}_i$, injected directly into the source antenna array. We assume that $\zeta^{\text{in}}$ is modulated at two distinct close-by frequencies $\omega_1$ and $\omega_2$, such that:
\begin{equation}\label{e1}
s^{\text{in}}_i=\zeta^{\text{in}}_i \bigg(\sin(\omega_1t)+ \sin(\omega_2t) \bigg).
  \end{equation}
The permittivity $\epsilon_r$ of the STMS is modulated with a depth $\delta_m$ and a phase $\phi$, at a frequency $\omega_m=|\omega_1-\omega_2|/2$, so that $\epsilon_r=\epsilon_s+\delta_m\cos(\omega_mt+\phi)$.  This choice of modulation frequency allows for the two input frequency to be efficiently mixed at the dominant Floquet harmonic $(\omega_1+\omega_2)/2$ (see \textcolor{blue}{Fig. \ref{Fig_1}(c)}). As we will now see, the reflection and transmission coefficients of Floquet Harmonics show a strongly non-linear dependency on the modulation phase, a key property that we will leverage to make the ELM very efficient.

To understand how time-Floquet systems can be used to induce large non-linear entanglement between the incident and reflected signals, let us consider the toy model of a generic two-port time-Floquet system, where  incident and reflected signals at  ports 1 and 2 are represented by their time-varying complex amplitudes $a_{1,2}(t)$ and $b_{1,2}(t)$. This model applies for each plane wave incident on our STMS, with transverse wave number $k$, on which the actual field can be decomposed. Assuming the modulation frequency $\omega_m$ to be much smaller that the operation frequency $\omega_k$ \cite{mousavi2014strong,salary2018electrically}, we can neglect dispersive effects and write the following instantaneous relation between the signals at each ports: \cite{salary2018electrically,mousavi2014strong,salary2018time}  
\begin{align}
 \begin{bmatrix}
a_1(t) \\
b_1(t)
\end{bmatrix}= \tilde\Psi(\omega_k,t)  \begin{bmatrix}
a_2(t) \\
b_2(t)
\end{bmatrix}, 
\end{align}
where $\tilde\Psi(\omega_k,t)$ is the transfer matrix at $\omega_k$, which varies slowly with time. Taking the Fourier transform of both sides yields
\begin{align}
 \begin{bmatrix}
A_1(\omega) \\
B_1(\omega)
\end{bmatrix}= \tilde\Psi(\omega_k,\omega)  *\begin{bmatrix}
A_2(\omega) \\
B_2(\omega)
\end{bmatrix}= \int \tilde\Psi(\omega_k,\omega-\omega')\begin{bmatrix}
A_2(\omega') \\
B_2(\omega')
\end{bmatrix} d\omega',
\end{align}
Since the scattering process into each Floquet harmonic component is linear, we can define the reflection and transmission coefficients into each harmonic as $    R_0(\omega_k+n\omega_m)={B_1(\omega_k+n\omega_m)}/{A_1(\omega_k)}$ and $     T_0(\omega_k+n\omega_m)={A_2(\omega_k+n\omega_m)}/{A_1(\omega_k)}$. A direct calculation shows that (see Supplementary for detail derivations): 
\begin{align}
  R_\phi(\omega_k+n\omega_m)=e^{in\phi} R_0(\omega_k+n\omega_m)  \\
  T_\phi(\omega_k+n\omega_m)=e^{in\phi} T_0(\omega_k+n\omega_m), 
\end{align}
where we have used the notation $R_\phi$ to highlight the dependency of the scattering coefficients on the modulation phase $\phi$. These equations imply that upon adding a  phase delay $\phi$ to the modulation, the generated frequency harmonic of
order $n$ will acquire a phase shift of $n\phi$, both for the forward and backward scattered plane waves. On the other hand, the amplitude of harmonic waves are constant when we alter the phase delay.
\begin{figure*}[t]
\centering
\includegraphics[scale=.6]{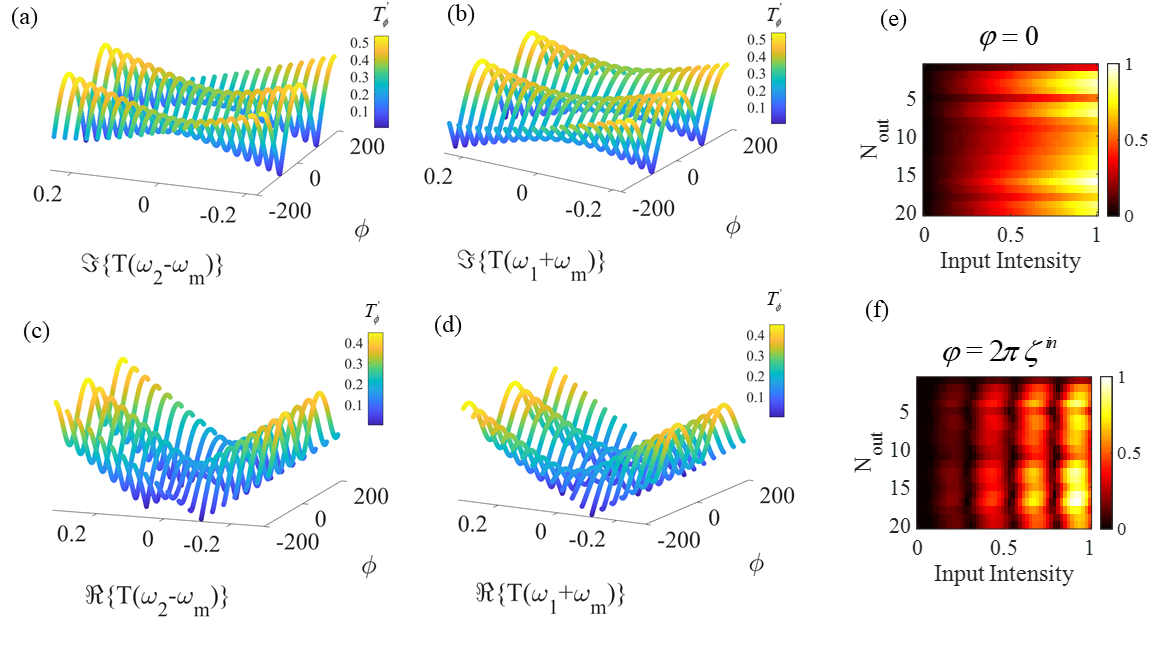}
\caption{\textbf{Non-linear Floquet entanglement} (a)-(d) Theoretical demonstration of the nonlinear dependency of the intensity $T'_{\phi}$ of the central Floquet harmonic ($\omega=(\omega_1+\omega_2)/2$) on both the modulation phase $\phi$ and the real or imaginary part of one of the generated harmonics. The results are based on Eq. (7). The fixed parameters for (a)-(d) are: 
(a): $T(\omega_1+\omega_m)=0.1-0.25i$ and $\Re{\{T(\omega_2-\omega_m)}\}=0.1$. (b): $T(\omega_2-\omega_m)=0.1-0.25i$ and $\Re{\{T(\omega_1+\omega_m)}\}=0.1$.  (c): $T(\omega_1+\omega_m)=0.1-0.05i$ and $\Im{\{T(\omega_2-\omega_m)}\}=0.05$. (d): $T(\omega_2-\omega_m)=0.1-0.05i$ and $\Im{\{T(\omega_1+\omega_m)}\}=0.05$.
(e) and (f)  The numerical demonstration of linear/nonlinear Floquet entanglement for the  central harmonic wave for different readout nodes in terms of input intensity, for static (e) and dynamic phase delays (f). }
\label{Fig_2}
\end{figure*}

Now, consider the superposition of two incident plane waves at frequencies $\omega_1$ and $\omega_2$. Recalling our choice of modulation frequency, namely  $\omega_m=|\omega_1-\omega_2|/2$,  we can write the reflection and transmission waves for all Floquet harmonic components of frequency $\omega_1+n\omega_m=\omega_2+m\omega_m$ by using the superposition principle: 
\begin{align}
  |R^{'}_{\phi}|=|e^{in\phi} R_0(\omega_1,\omega_1+n\omega_m)+e^{im\phi} R_0(\omega_2,\omega_2+m\omega_m)|  \\
  |T^{'}_{\phi}|=|e^{in\phi} T_0(\omega_1,\omega_1+n\omega_m)+e^{im\phi} T_0(\omega_2,\omega_2+m\omega_m)|, 
\end{align}
where $n$ and $m$ are the orders of the Floquet harmonics with respect to $\omega_1$ and $\omega_2$, respectively. A particular example is the harmonic located at the average frequency $\omega=(\omega_1+\omega_2)/2$, for which $n=1=-m$ (orange spectrum in \textcolor{blue}{Fig. \ref{Fig_1}(c)}. According to \textcolor{blue}{Eqs. (6) and (7)}, the relation between the modulation phase and the intensity of scattered harmonic fields is highly nonlinear.  In fact, we can control the amplitude of  the Floquet harmonics  only by changing the modulation phase. In order to have a nonlinear input-output mapping, we must therefore entangle the phase delay with the input input vector (i.e., $\phi=f(\zeta^{in})$), using for example a simple external signal mixer. In other words, the value of the modulation phase is directly determined by the value of the input vector, which is fixed when the system is excited, automatically making the scattering process a highly non-linear function of the input, regardless of the input power. This makes such time-Floquet non-linear entanglement highly advantageous in neuromorphic computing schemes.

To exemplify the strong nonlinear response of the proposed system, we plot the amplitude of the transmitted central harmonic ( $\omega=(\omega_1+\omega_2)/2$) as a function of various variables, including the phase delay phi. The results are displayed in \textcolor{blue}{Figs. \ref{Fig_2}(a)-(d)}. We fix one of the harmonics and plot $T'_{\phi}$ versus the modulation phase and the real or imaginary part of the other transmitted harmonics,$T(\omega_1+\omega_m)$ (or $T(\omega_2-\omega_m)$). As we can see in \textcolor{blue}{Figs. \ref{Fig_2}(a)-(d)}, we indeed obtain a complex non-linear semi-sinusoidal form for $T'_{\phi}$, upon altering the modulation phase. The dependency on the real or imaginary parts of the other transmitted harmonic is also always non-linear.

Next, we implement the entanglement with the input vector to demonstrate the complex nonlinear behavior of the Floquet system, using a full-wave finite-difference time-domain simulation of the set-up of \textcolor{blue}{Fig. \ref{Fig_1}(b)} (see Methods). We compute the intensity of the central harmonic with respect to the input intensity for two different scenarios: a static phase delay and a dynamic phase delay. In the first scenario, the phase delay is fixed and not dependent on the input ($\phi=0$), and as  shown in \textcolor{blue}{Fig. \ref{Fig_2}(e)}, the harmonic intensities are linear in term of input intensities. In the second scenario, the delay phase is a simple linear function of the input (i.e., $\phi=2\pi\zeta^{in})$). \textcolor{blue}{Fig. \ref{Fig_2}(f)} shows the complex nonlinear form of proposed system. 
The oscillating nonlinear mapping performed by the proposed system is completely different form any earlier approaches. As we will show, it is surprisingly effective in transforming the input data
space  to a nearly linearly separable output data space.

\subsection{Learning highly nonlinear functions}
\begin{figure*}[t]
\centering
\includegraphics[scale=.51]{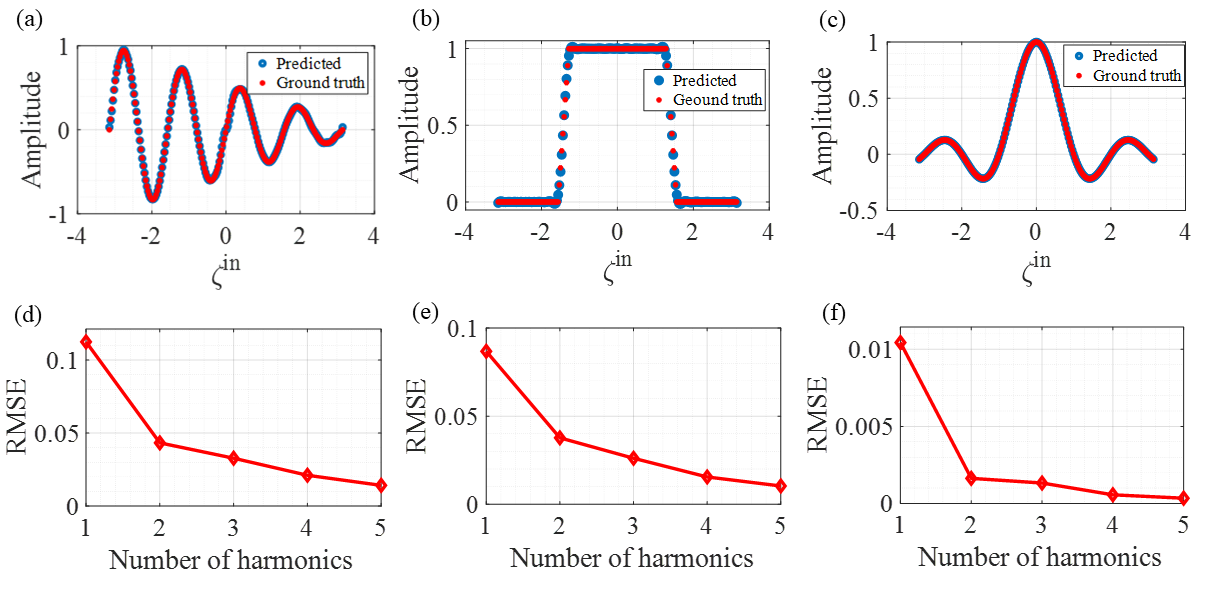}
\caption{\textbf{Floquet extreme learning for  highly nonlinear maps:} (a), (b), and (c) Comparison between ground truth and predicted values for three different nonlinear functions:  $y_1=\alpha\sin(4\pi\zeta^{in})(|\zeta^{in}|/\pi)$, and $y_2= \text{rect}(\zeta^{in})$ and $y_3 = \sin(\pi\zeta^{in})/(\pi\zeta^{in})$, respectively. (d), (e), and (f) The corresponding values of root-mean-square error (RMSE) upon increasing the numbers of involved Floquet harmonics at the readout nodes. }
\label{Fig_3}
\end{figure*}
 We now demonstrate the performance of the Floquet ELM by starting with  simple regression problems, on a dataset generated with  nonlinear relations. Such dataset is often used as a standard benchmark in machine learning since linear regression of a nonlinear function is impossible without a nonlinear transformation \cite{huang2006extreme, teugin2020scalable}. The input information ($\zeta^{in}$) is a set of randomly generated numbers between $-\pi$
to $\pi$ and the corresponding output labels ($y_i$) are generated according to non-linear functions, namely  $y_1=\alpha\sin(4\pi\zeta^{in})(|\zeta^{in}|/\pi)$,  $y_2= \text{rect}(\zeta^{in})$ (pulse function), and $y_3 = \sin(\pi\zeta^{in})/(\pi\zeta^{in})$. We use 1000 randomly generated samples, which lie in $[-\pi,\pi]$ to cover the entire characteristic behavior of the function. We map each input value to a vector by multiplying it with a fixed random 1D vector (mask), here of dimension $1\times 10$. In this task, we use 10 input nodes and readout nodes. By recording the intensity of the harmonics in the readout nodes of
many input values, a linear regression method is performed on the output data (see \textcolor{blue}{Figs. \ref{Fig_3}(a)-(c)}). A remarkable learning performance, with very low root-mean-squared error (RMSE) for all three nonlinear functions, is obtained.  Interestingly, in the proposed wave-based neural network with a  nonlinear time-Floquet layer, the multiple generated harmonic fields can be used to extend the dimension of the nonlinear mapping, and increasing their number improves the accuracy of classification/regression.
 This tendency is demonstrated in \textcolor{blue}{Figs. \ref{Fig_3}(d)-(f))}, which plots the RMSE versus the number of considered Floquet harmonics. This mechanism is a clear advantage of the Floquet ELM: by involving a higher number of scattered harmonics, we can improve the RMSE and enhance the accuracy of learning with no additional computational cost. Note that, in the last layer, a fast Fourier transform (FFT) is performed on the time-series output data obtained from the readout nodes, in order to compute the intensity of the different harmonic waves (see further details in Methods). However, performing FFT requires only a simple multiplication and does not impose a large computational overhead.

\subsection{Abalone dataset}
\begin{figure*}[t]
\centering
\includegraphics[scale=.51]{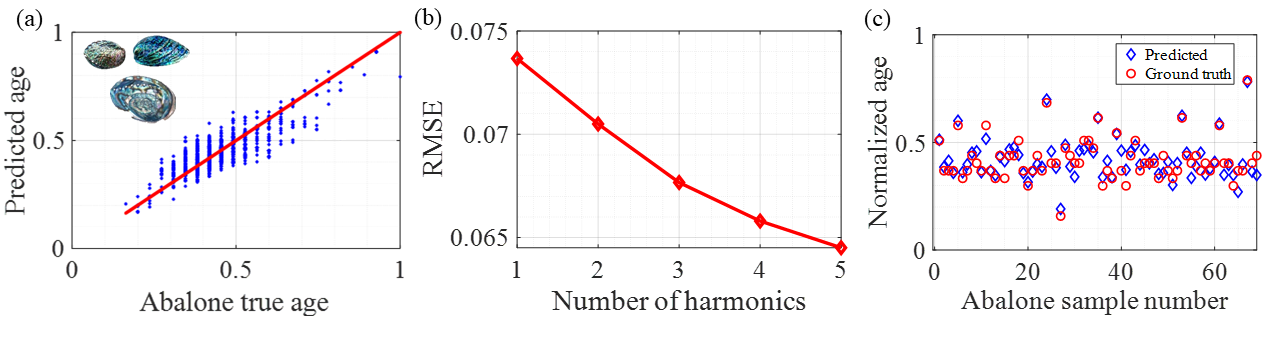}
\caption{ \textbf{Floquet extreme learning for multi-variable regression:} (a) and (b) Learning of the
abalone  dataset and corresponding RMSE for different numbers of considered harmonic waves, respectively. (c) Comparison
between predicted data (blue) and ground truth (red).}
\label{Fig_4}
\end{figure*}

In the previous section, we have used our Floquet ELM to learn nonlinear functions and its interpolation capability. However, interpolation is not always the relevant task, especially in complex inference problems. Therefore, we now move to a more challenging multivariable problem: the abalone dataset \cite{abolone}. This dataset is one of the most used benchmarks for machine learning, and concerns the classification of sea snails in terms of age and physical parameters.
It lists eight physical features of sea snails that can be used for the prediction of their age. 
To tackle this problem with our Floquet ELM, we encode the 8 physical features of sea snails on our input nodes, and consider 50 readout nodes to feed the decision layer, which performs linear regression. \textcolor{blue}{Fig. \ref{Fig_4}(a)} presents the true ages and the corresponding predictions; the figure indicates that the framework learns the ages of the abalone with remarkable accuracy. For a direct comparison, we plot the predicted values for 75 random input data (\textcolor{blue}{Fig. \ref{Fig_4}(c)}). The RMSE with respect to number of  harmonic waves are plotted in  \textcolor{blue}{Fig. \ref{Fig_4}(b)}. \textcolor{black}{A remarkable accuracy (RMSE=0.064) can be achieved by considering five generated harmonics. The achieved RMSE, is smaller than state-of-the-art reported values \cite{teugin2020scalable}.}

\subsection{Parallel image classifications}
 
\begin{figure*}[t]
\centering
\includegraphics[scale=.51]{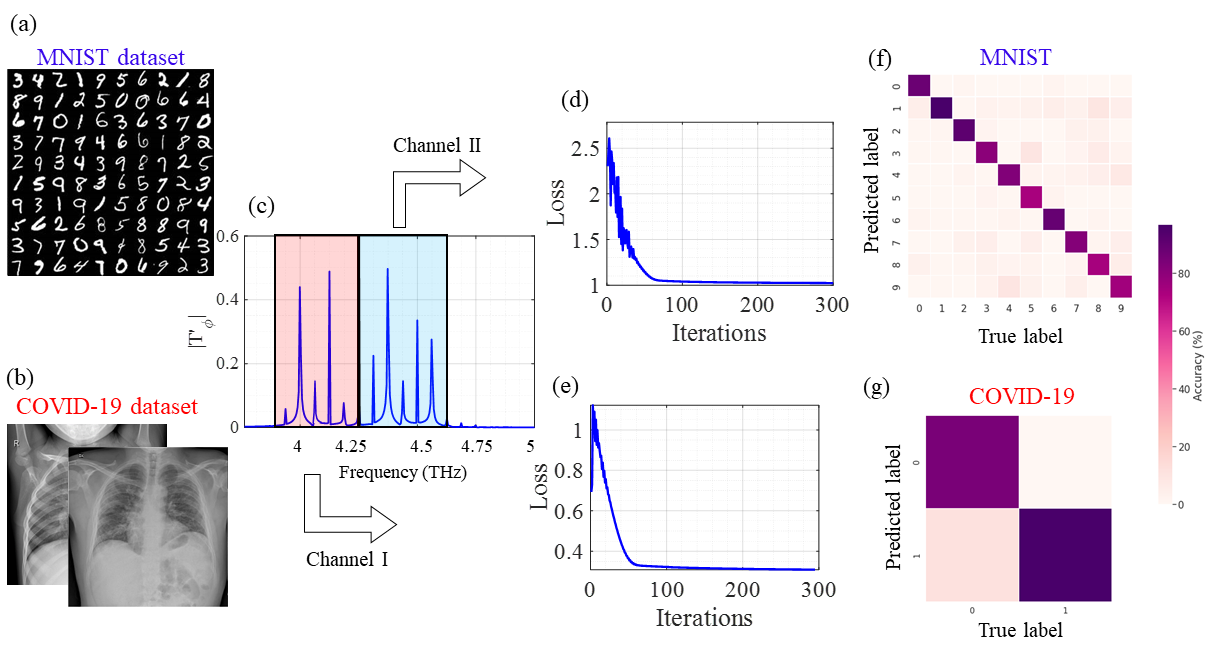}
\caption{\textbf{Floquet extreme learning for parallel image classification.} (a) and (b) Examples of realizations taken from the MNIST and COVID-19 X-ray datasets. (c) Simulated spectra of readout nodes for two different channels. (d) and (e) Evolution of the loss function for MNIST and COVID-19 classifications, respectively. (f) and (g) Corresponding  confusion matrices, for condition classification.}
\label{Fig_5}
\end{figure*}

Another remarkable feature of time-Floquet systems is that since the inputs are modulated at a certain carrier frequency, we can use several frequency bands and multiplex different signals to classify them simultaneously using the same system, and no additional cost in terms of power consumption. Let us now demonstrate this in a specific complex parallel classification task.  We examine the possibility to perform parallel image classification using two wavelength inputs. We use two distinct datasets: the MNIST dataset of handwritten digits \cite{mnist} and the COVID-19 X-ray images \cite{kaggle} (see \textcolor{blue}{Figs. \ref{Fig_5}(a) and (b)} ).  
We resize all of images into 10$\times$10 pixels, down-sampling them to decrease the number of input and readout nodes and the total size of our structure. In this task, we use 100 nodes to encode the images with the amplitude of the input waves.  
The MNIST data are encoded onto a (randomly selected) frequency range from 4 to 4.125 THz, and the COVID-19 data are encoded between 4.375 and 4.5 THz (see the red and blue frequency bands in \textcolor{blue}{Fig. \ref{Fig_5}(c)} ). In the output layer, we use Softmax regression to perform classifications (See Methods). 

The training results are shown in \textcolor{blue}{Figs. \ref{Fig_5}(d)-(g)} .  The observed test accuracies were $88.2\%$ for the COVID-19 and $85.3\%$ for the MNIST datasets.
These classification accuracies are competitive. For example, they are higher than the ones reported in reference \cite{nakajima2021scalable} (parallel image classification).  Also, the classification-accuracy results are comparable with other relevant works despite the decreasing the pixel sizes of all images \cite{midya2019reservoir}. 
In addition, this frequency multiplexing technique is the first demonstration of wave-based parallel task processing with extreme deep learning. This enables the use of wide-bandwidth as a computational resource, which significantly boosts the computation efficiency.   

\section{Nonlinear Time-Floquet-based RC system for autonomous forecasting chaotic time-series}

To show the high versatility of the proposed nonlinear time-Floquet neuromorphic computing system, we slightly modify it to implement a reservoir computing (RC) scheme. Consider an input vector $i(t)$ that is injected to a high-dimensional dynamical system called the reservoir. The reservoir is described by a vector $h(t)$  and the initial state of the reservoir is defined randomly. Let the $W_{res}$ matrix define the internal connections of the reservoir nodes and the $W_{in}$ matrix define the connections between the input and the reservoir nodes. Both matrices are initialized randomly and fixed during the whole RC training process. The state of each reservoir node is a scalar $h(t)$, which evolves according to the following recursive relation:
\begin{align}
    h(t+\tau)=F\bigg(w_{in}i(t)+w_{res}h(t)\bigg)
\end{align}
where $\tau$ is the discrete time step of the input and $F$ is a nonlinear function. From \textcolor{blue}{Eq. (8)}, we see that the reservoir is defined as a dynamical system provided with a unique memory property; namely, each consequent state of the reservoir contains some  information about its previous states and about the inputs injected until that time. In the training phase, the input $i(t)$ is fed to the reservoir, and the corresponding reservoir states are recursively calculated. The final step of the information processing is to perform a simple linear regression in order to minimize the RMSE that adjusts the $W_{out}$ weights. The output can be computed with $O(t)=W_{out}h(t)$. It should be noted that the output weights are the only parameters that are modified during the training.The input and reservoir weights are fixed throughout the whole computational process, and they are used to randomly project the input into a highdimensional space, which increases the linear separability of inputs.

\textcolor{black}{In our concrete scheme, we implement this memory using a feedback-loop,  and use the intensity of harmonic waves as a reservoir states. The reservoir computing in our scheme can be described by the following recursive relation: }

\begin{align}
    T'_{\phi}(t+\tau)=F\bigg(w_{in}i(t)+w_{res}v_h T'_{\phi}(t)\bigg)
\end{align}
 where F the nonlinear function describing our system, \textcolor{black}{ $v_h$ is a tunable parameter that selects one (or more)  harmonics as reservoir states and $T'_{\phi}$ is the intensity of transmission harmonic waves. }
In general, the RC and its different implementations have proven to be very successful for various tasks, such as spoken digits recognition, temporal
Exclusive OR task,  Mackey-Glass, or Nonlinear Autoregressive Moving Average time-series prediction \cite{brunner2018tutorial, bertschinger2004real}.

\begin{figure*}[t]
\centering
\includegraphics[scale=.51]{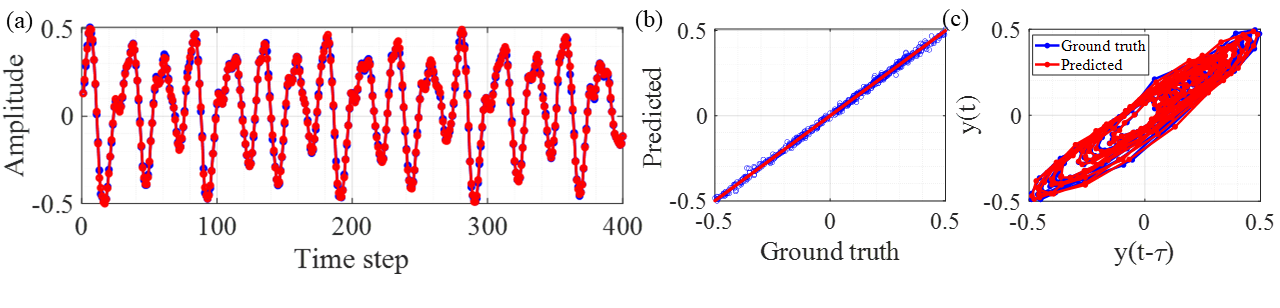}
\caption{ Floquet reservoir computing for forecasting the chaotic Mackey Glass time series. (a) Training results: The ground truth (blue) and the predicted output from the RC system (red) are plotted. (b) Corresponding results of linear regression. (c) Trace of time-series values in phase space, with respect to the previous time step for both ground truth and predicted values. }
\label{Fig_6}
\end{figure*}

We use the nonlinear time-Floquet RC for prediction of chaotic time series. Forecasting chaotic time series is a extremely difficult task due to the accumulation of quantitative difference between the ground truth and the predicted value in subsequent predictions, that lead to exponential errors at large times. Indeed, the positive Lyapunov exponent in chaotic systems leads to exponential growth for the separation of close trajectories, so that even small errors in prediction can quickly lead to divergence of the prediction from the ground truth \cite{moon2019temporal}.
We test  our system using the Mackey–Glass time series defined by \cite{mackey1977oscillation, moon2019temporal}.
\begin{align}
    \frac{dy}{dt}=\beta\frac{y(t-\tau)}{1+(y(t-\tau))^n}-\gamma y(t)
\end{align}
Unlike for deterministic equation, predicting such time-series for specific values of parameters is difficult and thus have been widely used as a benchmark for challenging forecasting tasks.
To obtain chaotic dynamics, here, we set the parameters $\beta=0.2$, $\gamma=0.1$, $\tau=18$, $n=10$.
During the training phase, as soon as the reservoir states  are  calculated, a simple linear regression is executed to adjust the $W_{out}$ weights such that their linear combination with the calculated reservoir states makes the actual output as close as possible to the next time step of the input. Finally, to automatically predict the future evolution of $i(t)$, we make a feedback-loop from the output to the input by replacing the next input $i(t+1)$  with the one-step prediction $W_{out}o(t)$, as is done in conventional RC. 
The ability of the proposed RC system in time-series prediction is tested using a reservoir with 75 nodes. We consider the middle harmonic as a reservoir state and input, $\zeta^{in}_n$, to feed our RC system for each interactions (\textcolor{blue}{Eq. (9)}). All of intensity harmonics, reservoir states, are then applied to the readout layer ( see Methods) to generate the predicted data for the next time step. \textcolor{blue}{Fig. \ref{Fig_6}} show the results obtained during training from the simulation. Excellent agreement between the target and the predicted value can be obtained, indicating that the trained readout  weights can correctly calculate the next time-step signal on the basis of the internal states of the reservoir. 
Further evidence of successful training can be found by examining the network performance in regression  and  phase space, as shown in \textcolor{blue}{Figs. \ref{Fig_6}(b)-(c)}, where an excellent agreement can again be observed .
\begin{figure*}[t]
\centering
\includegraphics[scale=.51]{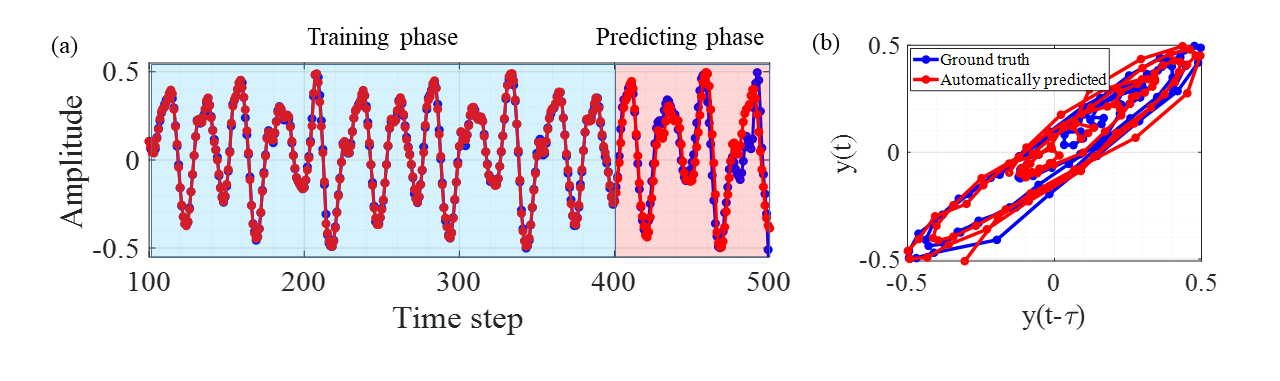}
\caption{\textbf{Autonomous forecasting of Mackey–Glass time series.} Training and forcasting results: The ground truth (blue) and the predicted output from the RC system (red)  for 100 next time steps are plotted.  (c) Trace of time-series values,   phase space, for predicting phase with respect to the previous time step.  }
\label{Fig_62}
\end{figure*}

The network is then used to forecast the time series autonomously. 
After training for 400 time steps, the output from the readout function, that is, the predicted data for the next time step, is then connected to the reservoir as the new input, and the system autonomously produces the forecasted time series continuously.
\textcolor{blue}{Fig. \ref{Fig_62}} show the results for autonomous time-series prediction using the proposed RC system. Afterwards, the autonomously generated output (from the 400th time step onwards) still matches very well the ground truth, showing the ability of the proposed RC system to autonomously forecast the chaotic system. After more than 70 time steps of autonomous prediction, the predicted signal starts to diverge from the correct value, which is unavoidable due to the chaotic nature of the series. Increasing the size of the reservoir further, by using more nodes and using more previous states  may  reduce the prediction error so that the length of accurate prediction can be  increased. Another solution for long-term forecasting without increasing the dimension of the system is utilizing a periodical update procedure as in Ref. \cite{moon2019temporal}.

\section{Conclusion}
In conclusion, we have shown how non-linear Floquet entanglement can be used to enable wave-based neuromorphic computing, by allowing for strong and tailored non-linear mapping to a higher dimensional space. Our nonlinear time-Floquet learning machine can  process information to compute complex tasks that are traditionally only tackled by slower, sophisticated, digital deep neural networks. In our benchmarks, the proposed  computing platform performs as well as its digital counterparts . With better energy efficiency in comparison to the previous proposals and a path to high scalability, our nonlinear time-Floquet system provides a novel path toward supercomputer-level optical computation.

\section{Methods}
\subsection{Numerical simulations}
\begin{figure*}[t]
\centering
\includegraphics[scale=.61]{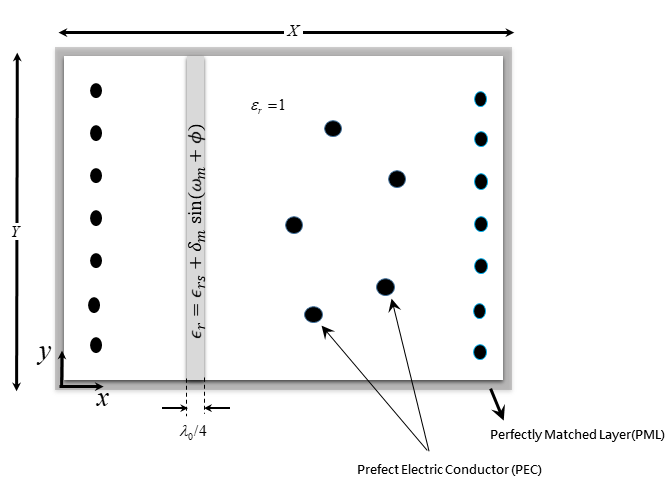}
\caption{\textbf{Layout of the setup.  } Black disks are an array of hard source (left side as an input array and right side as a receivers). The gray rectangle in the propagating substrate is a thin dielectric slab whose index of refraction is modulated at the frequency $\omega_m$ and five PEC sub-wavelength scatterers, randomly located in the propagating substrate. The gray rectangle represents an absorbing boundary layer. }
\label{Fig_8}
\end{figure*}

We use a two-dimensional finite-difference time-domain (FDTD) method for all simulations \cite{elsherbeni2016finite,kunz1993finite}. \textcolor{blue}{Fig. \ref{Fig_8} } shows the rectangular layout of the utilized setup. We set the parameters $\epsilon_{rs}=3$, $\delta_m=0.3$, $\omega_m=|\omega_1-\omega_2|/2$, $Y=8\lambda_0$, and $X=15\lambda_0$.
Also, the time window of simulation and the spatial window (time- and space-discretization factors) is set as a $d_t=d_{x,y}/(2C)$ and $d_{x,y}=\lambda_0/30$, respectively (C is speed of light). We use 10000 time-steps to ensure convergence has happened. \textcolor{black}{ In order to calculate the intensity of different harmonic waves, a simple fast Fourier transform (FFT) is performed. }

\subsection{Training of readout}
For learning nonlinear functions, Abolone dataset, and forecasting chaotic time series, we used a supervised learning algorithm, linear regression, to train the readout function.  The predicted output is compared with the ground truth, and the error is calculated and used to update the weights in the readout network following the linear regression learning rule. 

 To train the readout network, for classification task parallel image processing, we used the Python toolkit Keras, which provides a high-level application programming interface to access TensorFlow. A supervised learning algorithm, softmax regression, was used to train the readout network. A softmax function is used as the activation function of the readout network to calculate the probability corresponding to the different possible outputs. The cost is calculated following a categorical crossentropy. A standard gradient-based optimization method is used to minimize the cost function and train the output network. There are several ways of converting images into one-dimensional representations. For
simplicity, we used a flattened version of downsampled images as an output vector.

\textbf{Acknowledgements}:
A. Momeni and R. Fleury acknowledge funding from the Swiss National Science Foundation under the Eccellenza grant number 181232.



\bibliography{cas-refs}
\newpage
\section{\centering SUPPLEMENTARY MATERIAL}

The transfer matrix equation relating the amplitudes of the ﬁelds on opposite sides of the time-Floquet system (See Fig. 9) can be expressed in multiplicative form in time-domain for each excitation frequency as   
\begin{align}
 \begin{bmatrix}
a_1(t) \\
b_1(t)
\end{bmatrix}= \tilde\Psi(\omega_k,t)  \begin{bmatrix}
a_2(t) \\
b_2(t)
\end{bmatrix} 
\end{align}
where $\tilde\Psi(\omega_k,t)$ is the time-varying transfer matrix. The transfer matrix equation can be taken into angular frequency domain by taking the Fourier transform of both sides as:
\begin{align}
 \begin{bmatrix}
A_1(\omega) \\
B_1(\omega)
\end{bmatrix}= \tilde\Psi(\omega_k,\omega)  *\begin{bmatrix}
A_2(\omega) \\
B_2(\omega)
\end{bmatrix}= \int \tilde\Psi(\omega_k,\omega-\omega')\begin{bmatrix}
A_2(\omega') \\
B_2(\omega')
\end{bmatrix} d\omega'
\end{align}
Equation (12) implies that an input frequency $\omega_k$ will be converted to a spectrum of output frequencies. In the time-Floquet system, when the elements are varying in time with a periodic modulation having a modulation frequency of $\omega_m$, $\epsilon_r=\epsilon_s+\delta_m\cos(\omega_mt)$, the transfer
matrix is also periodic and can be expanded in form a Fourier series as $\tilde\Psi(t)=\sum_{n}\tilde\Psi^n(\omega_k)e^{in\omega_mt}$.  And the Fourier transform takes the following form:
\begin{align}
    \tilde\Psi(\omega)=\sum_{n}\tilde\Psi^n(\omega_k)\delta(\omega-n\omega_m)
\end{align}
By choosing $\omega=\omega_q=\omega_k+q\omega_m$ and $\omega'=\omega_p=\omega_k+p\omega_m$, q and p $\in\{...,-1,0,+1,...\}$,  and substituting equations (13) into (12), we will arrive at the following equations \cite{salary2018electrically}: 
\begin{align}\bigg\{
\begin{bmatrix}
A_1(\omega_q) \\
B_1(\omega_q)
\end{bmatrix}\bigg\}= \{\tilde\Psi^{q-p}(\omega_k)\} \bigg\{ \begin{bmatrix}
A_2(\omega_p) \\
B_2(\omega_p)
\end{bmatrix}\bigg\}
\end{align}
Now, let us consider adding a phase delay of $\phi$ to the sinusoidal modulation profile. Writing Equation (14) for the phase-delayed modulation and using $\tilde\Psi^{q-p}_\phi(\omega_k)=\text{exp}(i(q-p)\phi)\tilde\Psi^{q-p}_\phi(\omega_k)$, we have:
\begin{align}\bigg\{
\begin{bmatrix}
A_1^{(\phi)}(\omega_q) \\
B_1^{(\phi)}(\omega_q)
\end{bmatrix}\bigg\}= \{e^{(i(q-p)\phi)}\tilde\Psi^{q-p}(\omega_k)\} \bigg\{ \begin{bmatrix}
A_2^{(\phi)}(\omega_p) \\
0
\end{bmatrix}\bigg\}
\end{align}
By setting $B_2 (\omega_p) = 0$ for
all p's and $A1(\omega_q) = 0$ for all q's except $A_1(w0) = 1$, we can solve the equation to obtain the reflection and transmission coefficients of all generated frequency harmonics for a monochromatic excitation of $\omega_k$ incident
as: 
\begin{align}
  R_\phi(\omega_k+n\omega_m)=e^{in\phi} R_0(\omega_k+n\omega_m)  \\
  T_\phi(\omega_k+n\omega_m)=e^{in\phi} T_0(\omega_k+n\omega_m) 
\end{align}
where $    R_0(\omega_k+n\omega_m)={B_1(\omega_k+n\omega_m)}/{A_1(\omega_k)}$ and $     T_0(\omega_k+n\omega_m)={A_2(\omega_k+n\omega_m)}/{A_1(\omega_k)}$.

\begin{figure*}[t]
\centering
\includegraphics[scale=.81]{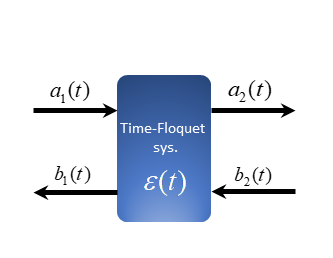}
\caption{The schematic of a generic two-port time-Floquet system. Incident and reflected signals at ports 1 and 2 are represented by their time-varying complex amplitudes $a_{1,2}(t)$, and $b_{1,2}(t)$.}
\label{Fig_9}
\end{figure*}
\end{document}